\documentclass[prd,notitlepage,longbibliography,nofootinbib,superscriptaddress,onecolumn,preprintnumbers]{revtex4-2}
\usepackage[utf8]{inputenc}

\usepackage{bm}
\usepackage{comment} 
\usepackage[colorlinks=true,urlcolor=blue,anchorcolor=black,citecolor=blue,linkcolor=red,filecolor=black,menucolor=black,pagecolor=black,linktocpage=true,pdfproducer=medialab,pdfa=true]{hyperref}
\usepackage{graphicx}
\usepackage{amsmath,latexsym,amssymb,mathrsfs,ascmac,mathtools}
\usepackage{multirow}
\usepackage{braket}
\usepackage{placeins}
\usepackage{float}
\usepackage{subcaption}
\usepackage{makecell}
\begin{document}

\title{Black Hole Entropy and Holographic Entanglement Entropy in DGP Brane Gravity}

\author{Takamasa Kanai}
\email{kanai@kochi-ct.ac.jp}

\affiliation{Department of Social Design Engineering,
National Institute of Technology (KOSEN), Kochi College,
200-1 Monobe Otsu, Nankoku, Kochi, 783-8508, Japan}

\begin{abstract}
We investigate the relation between gravitational entropy and holographic entanglement entropy in braneworld gravity with gravitational dynamics localized on the brane. We first consider the DGP brane model, in which an Einstein-Hilbert term is included in the brane action, and demonstrate that the gravitational entropy of a braneworld black hole agrees with the holographic entanglement entropy for general stationary spacetimes.

We then generalize the brane-localized Einstein-Hilbert term to higher-curvature gravitational theories. For Lovelock gravity and $f(R)$ gravity on the brane, we derive the corresponding expressions for gravitational entropy and holographic entanglement entropy and show that the two entropies also coincide for general stationary spacetimes.

These results establish the equivalence between gravitational entropy and holographic entanglement entropy in braneworld gravity for a broad class of gravitational theories and contribute to a deeper understanding of holography in braneworld settings.
\end{abstract}
\maketitle

\section{Introduction}

Gravitational entropy lies at the intersection of gravity, thermodynamics, and quantum theory. The Bekenstein--Hawking area law established that black hole horizons carry an entropy proportional to their area in Einstein gravity \cite{Hawking:1974rv,Hawking:1975vcx,Bekenstein:1972tm,Bekenstein:1973ur,Bekenstein:1974ax}. This simple geometric interpretation is modified in theories with higher-curvature interactions, where the entropy depends more generally on the curvature structure of the gravitational action \cite{Wald:1993nt,Jacobson:1993vj,Jacobson:1993xs}. The study of gravitational entropy therefore provides a way to explore aspects of gravitational dynamics that are not captured solely by the area of a horizon and may shed light on the underlying connection between gravity and quantum physics.

The holographic principle offers another important route to this connection through quantum entanglement. The AdS/CFT correspondence \cite{Maldacena:1997re,Gubser:1998bc,Witten:1998qj} provides a concrete realization of the idea that gravitational physics in a higher-dimensional spacetime can encode a quantum field theory in a lower-dimensional setting. One of the most striking manifestations of this correspondence is the geometric description of entanglement entropy. In Einstein gravity, the Ryu--Takayanagi prescription relates the entanglement entropy of a boundary region to the area of an appropriate codimension-two surface in the bulk \cite{Ryu:2006bv}, with the covariant formulation given by the Hubeny-Rangamani-Takayanagi prescription \cite{Hubeny:2007xt}. Such holographic constructions have also been developed in braneworld scenarios \cite{Gubser:1999vj,Giddings:2000mu,Shiromizu:2001jm,Shiromizu:2001ve,Nojiri:2000eb,Nojiri:2000gb,Hawking:2000kj,Koyama:2001rf,Kanno:2002iaa}, where the gravitational theory contains a lower-dimensional brane embedded in a higher-dimensional bulk. These settings consequently provide a natural framework for examining how gravitational entropy and quantum entanglement are related.

The relation between horizon entropy and holographic entanglement entropy has been investigated in a variety of braneworld settings. In particular, de Sitter braneworlds in the Randall--Sundrum II (RS II) model \cite{Randall:1999vf} provide an explicit realization of the correspondence between gravitational entropy and holographic entanglement entropy. For de Sitter branes embedded in an asymptotically anti-de Sitter bulk, the de Sitter entropy obtained from the Euclidean gravitational path integral was shown to agree with the corresponding holographic entanglement entropy \cite{Iwashita:2006zj}. This correspondence has also been extended to higher-curvature theories, including Gauss--Bonnet and Lovelock gravity \cite{Kushihara:2021fbr}. More recently, it has been shown that the equality between gravitational entropy and holographic entanglement entropy does not generally hold for stationary spacetimes in Ref.~\cite{Kanai:2026tnt}.

An especially interesting situation arises when gravity is not confined entirely to the bulk but also has its own dynamics on the brane. The Dvali-Gabadadze-Porrati (DGP) brane model \cite{Dvali:2000hr} is a representative example, with an Einstein--Hilbert term added to the brane action alongside the higher-dimensional bulk gravitational action. The brane-localized term contributes directly to the gravitational dynamics and consequently affects the entropy associated with black hole horizons on the brane. At the same time, the holographic entropy functional must account for the gravitational dynamics intrinsic to the brane. The comparison between these two entropy descriptions is therefore nontrivial. It is thus natural to investigate whether the non-equivalence found for general stationary spacetimes also arises in the DGP brane model and, if so, to identify the geometric condition governing the equality of the two entropies. 

In this work, we investigate the relation between gravitational entropy and holographic entanglement entropy for braneworld black holes in the DGP model. We first include an Einstein-Hilbert term in the brane action and derive expressions for the gravitational entropy and holographic entanglement entropy in this setup. We demonstrate that the two entropies coincide for general stationary spacetimes.

We then extend the analysis to more general gravitational theories localized on the brane. In particular, we replace the brane-localized Einstein-Hilbert term with Lovelock and $f(R)$ gravitational actions and investigate the corresponding relation between gravitational entropy and holographic entanglement entropy. Despite the higher-curvature structure of the brane action, we show that the two entropies coincide also in these theories. This demonstrates that the correspondence between gravitational entropy and holographic entanglement entropy holds for a broad class of gravitational theories localized on the brane, including Einstein, Lovelock, and $f(R)$ gravity.

Our results establish the general agreement between gravitational entropy and holographic entanglement entropy in braneworld gravity and show that this correspondence persists beyond Einstein gravity. These findings provide further insights into the relation between gravitational entropy and holographic entanglement entropy and contribute to a deeper understanding of holography in braneworld settings.

The remainder of this paper is organized as follows. In Sec.~\ref{sec:review}, we review the derivation of black hole entropy from the Euclidean gravitational action. In Sec.~\ref{sec:DGP}, we introduce the DGP brane model and investigate the relation between the black hole entropy obtained from the Euclidean gravitational action and the holographic entanglement entropy. We show that the two entropies coincide for general stationary braneworld black holes.

In Sec.~\ref{sec:lovelock}, we generalize the brane-localized Einstein-Hilbert term in the DGP model to Lovelock gravity and show that the equality between the two entropy prescriptions also holds for general stationary braneworld geometries. In Sec.~\ref{sec:fr}, we further extend the brane-localized gravitational action to $f(R)$ gravity and demonstrate that the two entropies likewise coincide for general stationary braneworld black holes.

Finally, in Sec.~\ref{sec:conclusion}, we discuss the implications of our results and their relation to braneworld holography and summarize our main results and discuss possible directions for future work.

For clarity, the geometrical quantities used throughout this paper are summarized in Table~\ref{tab:geometry}.
\begin{table}[htbp]
\centering
\caption{A summary of notation}
\label{tab:geometry}
\begin{tabular}{c|c|c|c|c|c}
\hline
  & metric $\&$ index  & curvature & covariant derivative & unit normal vector & surface type \\
\hline
$M_n$ & $g_{MN}$ & $R_{MNKL}[g]$ & $\nabla_M$ & Nothing & bulk\\
\hline
$\Sigma_{n}$ & $h_{ij}$ & $\mathcal{R}_{ijkl}[h]$ & $\bar{\nabla}_i$ &$u_M$ & Constant-time slice \\
\hline
$M_{n-1}$ & $q_{\mu\nu}$ &${}^{(n-1)}R_{\mu\nu\rho\sigma}[q]$ & $D_\mu$ & $n_{M}$ & Brane \\
\hline
$\Sigma_{n-1}$ & $\gamma_{XY}$ & ${}^{(n-2)}R_{XYZW}[\gamma]$ & $\tilde{\nabla}_X$ &$u_{M},n_{M}$ & Brane on a constant-time slice \\
\hline
$\Gamma^+\cup\Gamma^-$ & $G_{xy}$ & Not used & Not used & $u_{M},r_{M}$ & Cosmological horizon on a constant-time slice \\
\hline
$\partial\Gamma$ & $p_{AB}$ & ${}^{(n-3)}R_{ABCD}[p]$ & Not used & $u_{M},n_{M},r_{M}$ & Brane–horizon intersection on a constant-time slice \\
\hline
\end{tabular}
\end{table}

In this work, $H_{ij}$ and $K_{\mu\nu}$ denote the extrinsic curvatures of the constant-time hypersurface and the brane, respectively. The quantities ${}^{(n-2)}H_{XY}$ and ${}^{(n-2)}K_{XY}$ denote the extrinsic curvatures of the $(n-2)$-dimensional surface on the constant-time slice with respect to the normal vectors $u^M$ and $n^M$, respectively. Finally, ${}^{(n-3)}K_{AB}$ denotes the extrinsic curvature of $\partial\Gamma$ with respect to its normal vector $n^M$.

\begin{figure}[t]
    \centering
    \includegraphics[width=0.5\textwidth]{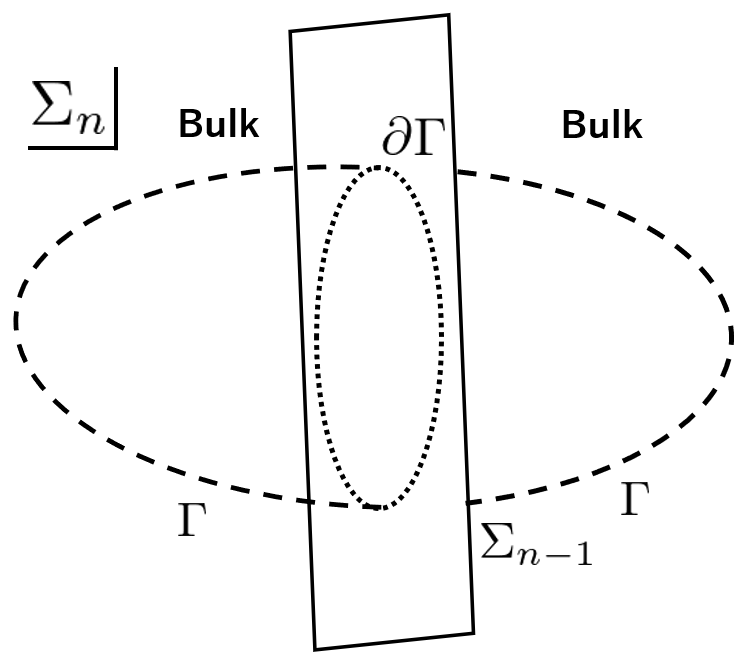}
    \caption{Schematic illustration of the black hole in DGP braneworld model at constant time.}
    \label{fig:spacetime}
\end{figure}

Figure~\ref{fig:spacetime} illustrates the spatial geometry on a constant-time slice. The central plane represents the $(n-1)$-dimensional brane, while the dashed curves represent the cosmological horizons $\Gamma$ in the $n$-dimensional bulk spacetime. The wavy curve represents the event horizon on the brane, denoted by $\partial\Gamma$.

We consider a DGP braneworld in which an $(n-1)$-dimensional brane is embedded in an $n$-dimensional bulk spacetime. In addition to the bulk Einstein-Hilbert term, the gravitational action contains an Einstein-Hilbert term localized on the brane $M_{n-1}$. We consider stationary braneworld black holes whose event horizon $\Gamma$ intersects the brane along a codimension-two surface $\partial\Gamma$, as illustrated in Fig.~\ref{fig:spacetime}. The corresponding bulk horizon extends away from the brane into the bulk spacetime, providing the geometric setting for comparing the gravitational black hole entropy with the holographic entanglement entropy. We further extend the analysis by considering Lovelock gravity and $f(R)$ gravity localized on the brane, and examine whether the relation between gravitational entropy and holographic entanglement entropy persists in these higher-curvature theories.

In this paper, we use square brackets to indicate complete antisymmetrization over the enclosed indices. For a rank-$n$ tensor $A_{a_1\cdots a_n}$, we define
\begin{align}
A_{[a_1a_2\cdots a_n]}:=\frac{1}{n!}\sum_{\sigma\in S_n}\operatorname{sgn}(\sigma)A_{a_{\sigma(1)}a_{\sigma(2)}\cdots a_{\sigma(n)}},
\end{align}
where $S_n$ is the set of all permutations of $n$ indices, and $\operatorname{sgn}(\sigma)$ is the sign of the permutation $\sigma$, taking the value $+1$ for even permutations and $-1$ for odd permutations.

\section{Review of Black Hole Entropy from the Euclidean Gravitational Action}
\label{sec:review}

We begin by reviewing the derivation of gravitational entropy from the Euclidean gravitational action. Consider an $n$-dimensional stationary spacetime containing a black hole horizon, and analytically continue the time coordinate to the Euclidean sector. We choose the time coordinate such that the corresponding time-evolution vector is proportional to the null generator of the horizon when evaluated on the horizon. This choice does not restrict the generality of our analysis. We further assume that the Euclidean time $\tau=it$ is periodic with period $\beta_0$.

The Euclideanized metric can then be expressed in the ADM form
\begin{align}
ds_E^2
&=N^2d\tau^2+h_{ij}\left(dx^i-iN^id\tau\right)
\left(dx^j-iN^jd\tau\right),
\end{align}
where $N$ and $N^i$ are the lapse and shift, respectively, and $h_{ij}$ is the induced metric on a constant-$\tau$ hypersurface. In what follows, the subscript $E$ indicates quantities defined in the Euclideanized spacetime.

We assume that the location of the Euclideanized horizon,
\begin{align}
\mathcal{H}_E:=\Gamma_E\times S^1,
\end{align}
is characterized by $N(x)=0$. We also choose the spatial shift such that
\begin{align}
N^i=0
\end{align}
on $\mathcal{H}_E$. In the vicinity of the horizon, the metric then takes the form
\begin{align}
ds_E^2&\approx N^2d\tau^2+h*{ij}dx^idx^j,
\end{align}
as discussed in Ref.~\cite{Brown:1993ke}. Since $N(x)$ is independent of the Euclidean time $\tau$, the orbit of $\tau$ on each surface of constant $N$ is an $S^1$. At the horizon, this $S^1$ orbit shrinks to zero size. To regulate the resulting singular behavior, we introduce a cutoff surface defined by $N(x)=\epsilon$ and regard
\begin{align}
\mathcal{H}_{E,\epsilon}:=\Gamma_{E,\epsilon}\times S^1
\end{align}
as a virtual boundary. The spacetime is then regarded as being bounded by the outer boundary and $\mathcal{H}_{E,\epsilon}$. No boundary condition is imposed on the virtual boundary $\mathcal{H}_{E,\epsilon}$. After completing the calculation, we take the limit $\epsilon\rightarrow0$.

In the neighborhood of $\mathcal{H}_E$, the metric can be expressed as
\begin{align}
ds_E^2&\approx N^2d\tau^2+\alpha^2dN^2+\cdots \notag\\
&=\rho^2d\left(\frac{\tau}{\alpha}\right)^2+d\rho^2+\cdots,
\end{align}
where
\begin{align}
\alpha^2&=g_{NN}\big|_{\Gamma_{E,\epsilon}},
\end{align}
and we have introduced the radial coordinate $\rho=\alpha N$. On $\Gamma_{E,\epsilon}$, we have
\begin{align}
\alpha^{-1}&=\lim_{\epsilon\rightarrow0}r^M\partial_MN,
\end{align}
where $r^M$ is the unit normal vector to $\Gamma_{E,\epsilon}\times S^1$, pointing inward from the boundary $\mathcal{H}_{E,\epsilon}$. The absence of a conical singularity therefore requires
\begin{align}
\label{conical condition BH}
\lim_{\epsilon\rightarrow0}\beta_0 r^M\partial_MN&=2\pi,\qquad\text{on }\Gamma_{E,\epsilon},
\end{align}
as discussed in Refs.~\cite{Solodukhin:1994yz,Fursaev:1995ef}. Under this condition, the deficit angle vanishes and the Euclideanized geometry is regular at the horizon.

We next rewrite the gravitational action in canonical form. For this purpose, we perform the ADM decomposition with respect to the Euclidean time $\tau$. The subscript $\epsilon$ denotes quantities in the spacetime with a cutoff surface $N(x)=\epsilon$. The canonical form of the action after subtracting the boundary term at infinity is written as
\begin{align}
I_{E,\epsilon}&=\beta_0\int_{\Sigma_{n,E,\epsilon}} d^{n-1}x\left[-i\pi^{ij}\partial_\tau h_{ij}+N\mathcal{H}+N^i\mathcal{H}_i\right]+\beta_0\int_{\mathcal{H}_{E,\epsilon}}d^{n-2}x\sqrt{G}\mathcal{L}_{\mathcal{H}},
\end{align}
where $\pi^{ij}$ is the momentum conjugate to $h_{ij}$, while $\mathcal{H}$ and $\mathcal{H}_i$ denote the Hamiltonian and momentum constraints, respectively, and $\mathcal{L}_{\mathcal{H}}$ denotes the boundary Lagrangian density associated with the virtual boundary $\mathcal{H}_{E,\epsilon}$. In this work, we denote by $I_{E,\epsilon}$ the Euclidean gravitational action with the boundary contribution at infinity subtracted.

For a stationary spacetime, the canonical variables are independent of the Euclidean time. Moreover, when the bulk spacetime satisfies the equations of motion, the constraints satisfy
\begin{align}
\mathcal{H}=\mathcal{H}_i=0.
\end{align}
Consequently, the bulk contribution to the canonical action vanishes on shell, and the Euclidean action is determined by the boundary contributions. In particular, the contribution associated with the virtual boundary $\mathcal{H}_{E,\epsilon}$ contains the information relevant for the black hole entropy.

We work in the canonical ensemble, in which the Euclidean time period $\beta_0$ is fixed and determines the inverse temperature. The gravitational entropy is then extracted from the Euclidean gravitational action by isolating the contribution associated with the black hole horizon. In the present formulation, this gives
\begin{align}
S_{\rm BH}&=-\lim_{\epsilon\rightarrow0}I_{E,\epsilon}(\beta_0).
\end{align}
This method for deriving the entropy was first introduced in Ref.~\cite{Gibbons:1976ue}, while its thermodynamic interpretation was subsequently clarified in Refs.~\cite{York:1986it,Braden:1990hw,Brown:1992bq}. Using the canonical form of the action and taking the limit $\epsilon\rightarrow0$, the contribution from the virtual boundary gives the gravitational entropy associated with the black hole horizon. This procedure provides a direct evaluation of the black hole entropy from the Euclidean gravitational action without imposing a boundary condition on the virtual horizon boundary.

\subsection{Holographic Entanglement Entropy from the Replica Method}

We briefly review the derivation of holographic entanglement entropy using the replica method. Let $A$ be a spatial subregion of the boundary theory and $\rho_A$ its reduced density matrix. The entanglement entropy is defined by
\begin{align}
S_{\rm EE}=-\mathrm{Tr}(\rho_A\log\rho_A).
\end{align}
The replica method provides a systematic way to compute the entanglement entropy by introducing $q$ replicas of the reduced density matrix $\rho_A$. One first considers the quantity
\begin{align}
\mathrm{Tr}(\rho_A^q),
\end{align}
which defines the Rényi entropy as
\begin{align}
S_q=\frac{1}{1-q}\log\mathrm{Tr}(\rho_A^q).
\end{align}
The entanglement entropy is then obtained by analytically continuing the replica number $q$ and taking the limit $q\to1$,
\begin{align}
S_{\rm EE}
=\lim_{q\to1}S_q
=-\left.\partial_q\log\mathrm{Tr}(\rho_A^q)\right|_{q=1}.
\end{align}
Thus, the replica method reduces the calculation of the entanglement entropy to the evaluation of $\mathrm{Tr}(\rho_A^q)$ for integer $q$ followed by an analytic continuation to $q\to1$.

In the Euclidean formulation, $\mathrm{Tr}(\rho_A^q)$ is represented by a path integral over a $q$-fold replicated geometry $\mathcal{M}_q$, obtained by cyclically gluing $q$ copies of the original geometry along the entangling surface. In a holographic theory, the corresponding partition function is evaluated semiclassically as
\begin{align}
\mathrm{Tr}(\rho_A^q)=\dfrac{Z_q}{(Z_1)^q}\simeq \dfrac{e^{-I_E[\mathcal{M}_q]}}{(e^{-I_E[\mathcal{M}_1]})^q},
\end{align}
where $I_E$ denotes the Euclidean gravitational action. The replicated bulk geometry possesses a $\mathbb{Z}_q$ replica symmetry, whose fixed-point locus is a codimension-two surface $\Gamma_q$. For $q\neq1$, this surface is associated with a conical singularity.

Taking the limit $q\rightarrow1$, the conical singularity disappears, and $\Gamma_q$ approaches a smooth codimension-two extremal surface $\Gamma_E$. For Einstein gravity, the contribution of the conical singularity to the gravitational action gives the area of this surface, leading to the Ryu--Takayanagi formula,
\begin{align}
S_{\rm HEE}=\frac{1}{4G_n}\int_{\Gamma_E}d^{n-2}x,\sqrt{G}.
\end{align}
Here, $G$ denotes the determinant of the induced metric on $\Gamma_E$. The extremality of $\Gamma_E$ follows from the gravitational equations of motion in the replica limit. In time-dependent or stationary settings, $\Gamma_E$ should be understood as a covariant extremal surface.

The same replica construction applies when the gravitational action contains terms localized on a brane. In the DGP brane model, the gravitational action contains both a bulk Einstein--Hilbert term and a brane-localized Einstein--Hilbert term. Consequently, the replica calculation receives contributions from both the bulk extremal surface and its intersection with the brane. Denoting the bulk extremal surface by $\Gamma_{E,\epsilon}$ and its intersection with the brane by $\partial\Gamma_{E,\epsilon}$, the holographic entanglement entropy is given by
\begin{align}
S_{\rm HEE}=\frac{1}{4G_n}\int_{\Gamma_{E}}d^{n-2}x,\sqrt{G}+\frac{1}{4G_{n-1}}\int_{\partial\Gamma_{E}}d^{n-3}x,\sqrt{p}.
\label{HEE EH}
\end{align}
Here, $p$ denotes the determinant of the induced metric on $\partial\Gamma_{E,\epsilon}$. The second term originates from the brane-localized Einstein-Hilbert term and represents the contribution of the brane to the holographic entanglement entropy.

For a more general gravitational theory, the holographic entanglement entropy is likewise obtained from the replica construction \cite{Lewkowycz:2013nqa,Dong:2013qoa,Camps:2013zua}. In particular, it is given by
\begin{align}
S_{\rm HEE}=\ \lim_{q\rightarrow1}\partial_q\left[I_E[\mathcal{M}_q]-q I_E[\mathcal{M}_1]\right],
\label{replica formula}
\end{align}
where $I_E[\mathcal{M}_q]$ is the Euclidean gravitational action evaluated on the replicated geometry. In the replica method, one constructs a $q$-fold replica geometry by taking $q$ copies of the original spacetime and cyclically gluing them along the entangling surface. The resulting geometry has a conical defect (or excess angle) around the entangling surface, whose contribution determines the entanglement entropy in the limit $q\to1$. This formula applies generally to gravitational theories with higher-curvature or higher-derivative terms. Evaluating the replica action in the $q\rightarrow1$ limit yields the corresponding gravitational entropy functional on the codimension-two extremal surface. For Einstein gravity, this reduces to the area functional, while for Lovelock gravity it gives the Jacobson-Myers functional \cite{Hung:2011xb}.

\section{Gravitational entropy in DGP model and Holographic entanglement entropy}
\label{sec:DGP}

In this section, we investigate the relation between black hole entropy and holographic entanglement entropy in the Dvali-Gabadadze-Porrati (DGP) brane model \cite{Dvali:2000hr}. Unlike the Randall-Sundrum (RS) model \cite{Randall:1999vf}, the DGP model contains an Einstein-Hilbert term localized on the brane, giving rise to gravitational dynamics intrinsic to the brane. This additional gravitational sector allows us to examine how brane-localized gravity affects the relation between gravitational entropy and holographic entanglement entropy. In particular, studying this relation in the DGP model allows us to examine whether the entropy correspondence in braneworld holography depends on the gravitational dynamics localized on the brane.

We consider the DGP model, in which the brane-localized Einstein-Hilbert term is supplemented by the higher-dimensional bulk gravitational action. We first introduce the corresponding action and then investigate the black hole entropy obtained from the Euclidean gravitational action and its relation to the holographic entanglement entropy.
\begin{align}
I=\frac{1}{16\pi G_n}\int_{M_n} d^nx\sqrt{-g}\left(R-2\Lambda\right)+\int_{M_{n-1}} d^{n-1}x\sqrt{-q}\left(-\sigma+\frac{1}{16\pi G_{n-1}}{}^{(n-1)}R+\frac{1}{8\pi G_n}[K]^-\right)+I_{\infty},
\end{align}
where  $I_{\infty}$ is the boundary terms at infinity, $g_{MN}$ and $q_{\mu\nu}$ denote the metrics of the $n$-dimensional bulk spacetime and the $(n-1)$-dimensional brane, respectively. The scalar $K$ denotes the trace of the extrinsic curvature of the brane $K_{\mu\nu}$, which appears in the Gibbons--Hawking boundary term. The quantities $R$ and ${}^{(n-1)}R$ are the Ricci scalars constructed from the bulk metric $g_{MN}$ and the induced metric $q_{\mu\nu}$ on the brane, respectively, while $G_n$ and $G_{n-1}$ denote the gravitational coupling constants in the bulk and on the brane, respectively. We assume that the brane $M_{n-1}$ is located at $y=0$ and introduce Gaussian normal coordinates $(y,x^\mu)$ in a neighborhood of the brane. The discontinuity of the trace of the extrinsic curvature across the brane is then defined as
\begin{align}
[K]^- := \lim_{y\rightarrow +0}K-\lim_{y\rightarrow -0}K.
\end{align}

Variation of the action gives
\begin{align}
\delta I=&\frac{1}{16\pi G_n}\int_{M_n} d^nx\sqrt{-g}\left(R_{MN}-\frac{1}{2}g_{MN}R+\Lambda g_{MN}\right)\notag\\
&+\frac{1}{16\pi G_{n-1}}\int_{M_{n-1}} d^{n-1}x\sqrt{-q}\left({}^{(n-1)}R_{\mu\nu}-\frac{1}{2}q_{\mu\nu}{}^{(n-1)}R-8\pi G_{n-1}{}^{(n-1)}T_{\mu\nu}\right),
\end{align}
where ${}^{(n-1)}T_{\mu\nu}$ is the energy momentum tensor o  the brane, given by
\begin{align}
{}^{(n-1)}T_{\mu\nu}=-\sigma q_{\mu\nu}-\frac{1}{8\pi G_n}[K_{\mu\nu}-q_{\mu\nu}K]^-.
\label{T}
\end{align}
We first derive the effective gravitational equation on the brane by varying the DGP action with respect to the bulk and induced metrics. The variation with respect to the bulk metric yields the $n$-dimensional vacuum Einstein equation,
\begin{align}
G_{MN}+\Lambda g_{MN}=0,
\label{bulk eq}
\end{align}
while the variation with respect to the induced metric gives the brane equation
\begin{align}
{}^{(n-1)}G_{\mu\nu}=8\pi G_{n-1}{}^{(n-1)}T_{\mu\nu}.
\label{brane eq}
\end{align}

To obtain a closed set of equations on the brane, we subsequently employ the Gauss equation, which relates the $(n-1)$-dimensional Riemann tensor to the $n$-dimensional Riemann tensor and the extrinsic curvature as
\begin{align}
{}^{(n-1)}R_{\mu\rho\nu\sigma}
={}&R_{MNKL}q^M{}_\mu q^N{}_\rho q^K{}_\nu q^L{}_\sigma+K_{\mu\nu}K_{\rho\sigma}-K_{\mu\sigma}K_{\rho\nu}.
\end{align}
Using the bulk vacuum equation, $R_{MN}=\frac{2\Lambda}{n-2}g_{MN}$, together with the fact that the $n$-dimensional Riemann tensor reduces to the Weyl tensor, this relation can be written as
\begin{align}
{}^{(n-1)}R_{\mu\nu}=\frac{2\Lambda}{n-2}q_{\mu\nu}-E_{\mu\nu}+K K_{\mu\nu}-K_{\mu\lambda}K^\lambda{}_\nu,
\end{align}
where
\begin{align}
E_{\mu\nu}=C_{MNKL}q^M{}_\mu q^K{}_\nu n^N n^L,
\end{align}
is the electric part of the bulk Weyl tensor. Since $E_{\mu\nu}$ generally depends on the bulk geometry, it represents a non-local contribution from the $n$-dimensional brane perspective.

Taking the trace of the contracted Gauss equation gives
\begin{align}
{}^{(n-1)}R&=\frac{2(n-1)}{n-2}\Lambda+K^2-K_{\mu\nu}K^{\mu\nu},
\end{align}
and hence the $(n-1)$-dimensional Einstein tensor is given by
\begin{align}
{}^{(n-1)}G_{\mu\nu}={}&\frac{3-n}{n-2}\Lambda q_{\mu\nu}-E_{\mu\nu}+K K_{\mu\nu}-K_{\mu\lambda}K^\lambda_{\ \nu}-\frac12 q_{\mu\nu}\left(K^2-K_{\rho\sigma}K^{\rho\sigma}\right).
\end{align}
The resulting effective Einstein equations on the brane were first derived in Ref.~\cite{Shiromizu:1999wj}. Thus, the gravitational dynamics of the system are governed by the coupled bulk and brane equations of motion, Eqs.~(\ref{bulk eq}) and (\ref{brane eq}), respectively.

We now turn to the calculation of the gravitational entropy. For this purpose, we rewrite the gravitational action in canonical form, which is particularly convenient for evaluating the entropy. The resulting canonical action is given by
\begin{align}
I=&\int_{M_n} d^nx\left(\pi^{ij}\dot{h}_{ij}-N\mathcal{H}-N^i\mathcal{H}_i-\frac{\sqrt{-g}}{8\pi G_n}\nabla_M(u^N\nabla_Nu^M-u^MH)-2\sqrt{h}\bar{\nabla}_i\left(\frac{N_j\pi^{ij}}{\sqrt{h}}\right)\right)\notag\\
&+\int_{M_{n-1}} d^{n-1}x\sqrt{-q}\Biggl(\frac{\hat{\pi}^{XY}}{\sqrt{-q}}\dot{\gamma}_{XY}-\sigma+\frac{1}{8\pi G_{n-1}}{}^{(n-1)}G_{\mu\nu}u^\mu u^\nu+\frac{1}{8\pi G_n}[K]^-+\frac{1}{8\pi G_{n-1}}{}^{(n-1)}G_{\mu\nu}u^\mu h^\nu_{\ i}\frac{N^i}{N}\notag\\
&\qquad\qquad\qquad\qquad-\frac{1}{8\pi G_{n-1}}D_\mu(u^\nu D_\nu u^\mu-u^\mu {}^{(n-2)}H)-\frac{2}{N}\tilde{\nabla}_X\left(\frac{N_Y\hat{\pi}^{XY}}{\sqrt{\gamma}}\right)\Biggr)+I_{\infty},
\end{align}
where $H$ is the trace of the extrinsic curvature $H_{ij}$ of the constant-time hypersurface, $\bar{\nabla}_i$ denotes the covariant derivative on the constant-time hypersurface, $D_\mu$ denotes the covariant derivative on the brane, and $\tilde{\nabla}_X$ denotes the covariant derivative on the constant-time hypersurface of the brane. The quantity $\pi_{ij}$ is the conjugate momentum given by
\begin{align}
\pi_{ij}=\frac{\sqrt{h}}{16\pi G_n}\left(H_{ij}-h_{ij}H\right),
\end{align}
and $\hat{\pi}_{XY}$ is defined as
\begin{align}
\hat{\pi}_{XY}=\frac{\sqrt{\gamma}}{16\pi G_{n-1}}\left({}^{(n-2)}H_{XY}-\gamma_{XY}{}^{(n-2)}H\right).
\end{align}
Here, ${}^{(n-2)}H_{XY}$ denotes the extrinsic curvature of the codimension-two surface on the constant-time brane hypersurface with respect to the normal vector $u^M$, while ${}^{(n-2)}H$ denotes its trace.

The Hamiltonian and momentum constraints are defined by
\begin{align}
\mathcal{H}&=\frac{16\pi G_n}{\sqrt{h}}\left(\pi_{ij}\pi^{ij}-\frac{({\rm tr}\pi)^2}{n-2}\right)-\frac{\sqrt{h}}{16\pi G_n}(\mathcal{R}-2\Lambda),\\
\mathcal{H}_i&=-2\sqrt{h}h_{ik}\bar{\nabla}_j\pi^{jk},
\end{align}
respectively.

Rewriting the action in canonical form and subsequently performing the Euclidean continuation, we obtain the Euclidean action in the form:
\begin{align}
I_{E,\epsilon}=&\frac{1}{16\pi G_n}\int_{M_{n,E,\epsilon}} d^nx\sqrt{g}\left(-R+2\Lambda\right)+\int_{M_{n-1,E,\epsilon}} d^{n-1}x\sqrt{q}\left(\sigma-\frac{1}{16\pi G_{n-1}}{}^{(n-1)}R-\frac{1}{8\pi G_n}[K]^-\right)\notag\\
=&\int_{M_{n,E,\epsilon}} d^nx\left(-i\pi^{ij}\dot{h}_{ij}+N\mathcal{H}+N^i\mathcal{H}_i\right)\notag\\
&+\int_{M_{n-1,E,\epsilon}} d^{n-1}x\sqrt{q}\left(\sigma-\frac{1}{8\pi G_{n-1}}{}^{(n-1)}G_{\mu\nu}u^\mu u^\nu-\frac{1}{8\pi G_n}[{}^{(n-2)}K]^--\frac{1}{8\pi G_{n-1}}{}^{(n-1)}G_{\mu\nu}u^\mu \gamma^\nu_{\ X}\frac{N^X}{N}+\frac{2n_iN_j\pi^{ij}}{N\sqrt{h}}\right)\notag\\
&-\beta_0\int_{\Gamma_{E,\epsilon}}d^{n-2}x\sqrt{G}\left(\frac{r^i\partial_iN}{8\pi G_n}-\frac{2}{\sqrt{h}}r_iN_j\pi^{ij}\right)-\beta_0\int_{\partial\Gamma_{E,\epsilon}}d^{n-3}x\sqrt{p}\left(\frac{r^X\partial_XN}{8\pi G_{n-1}}-\frac{2}{\sqrt{h}}\frac{G_n}{G_{n-1}}r_XN_Y\gamma^X_{\ Z}\gamma^Y_{\ W}\pi^{ZW}\right)+I'_{\infty,E,\epsilon}\notag\\
=&\frac{\beta_0}{8\pi G_{n}}\int_{\Sigma_{n-1,E,\epsilon}} d^{n-2}x\sqrt{\gamma}[K_{\mu\nu}u^\mu \gamma^\nu_{\ X}N^X]^--\frac{\beta_0}{8\pi G_n}\int_{\Gamma_{E,\epsilon}}d^{n-2}x\sqrt{G}r^i\partial_iN-\frac{\beta_0}{8\pi G_{n-1}}\int_{\partial\Gamma_{E,\epsilon}}d^{n-3}x\sqrt{p}r^X\partial_XN+I'_{\infty,E,\epsilon}.
\end{align}
Here, $I'_{\infty,E,\epsilon}$ denotes the sum of the bulk total derivative term and $I_{\infty,E,\epsilon}$. Throughout this paper, this notation is used irrespective of the choice of gravitational theory on the brane. In the second equality, we have used the relations
\begin{align}
K_{\mu\nu}u^\mu u^\nu+K
&=(g^{MN}+u^Mu^N-n^Mn^N)\nabla_Mn_N
={}^{(n-2)}K,\\
u^Mn_M&=u^Mr_M=0,
\label{relation eq1}
\end{align}
whereas, in the third equality, we have used the stationarity of the spacetime, the Hamiltonian and momentum constraints, $\dot{h}_{ij}=\mathcal{H}=\mathcal{H}_i=0$, which follow from the bulk equations of motion, Eq.~(\ref{bulk eq}), and the brane equations of motion, Eq.~(\ref{brane eq}), together with the condition $N^i=0$ on the horizon. In the last equality, we have used the $(u,i)$ components of the equations of motion.

\begin{align}
{}^{(n-1)}G_{\mu\nu}u^\mu\gamma^\nu_{\ X}
&=-\frac{G_{n-1}}{G_n}[K_{\mu\nu}u^\mu\gamma^\nu_{\ X}]^-,\\
\frac{2}{\sqrt{h}}n_iN_j\pi^{ij}
&=\frac{1}{8\pi G_n}n_iN_j\left(H^{ij}-h^{ij}H\right)\notag\\
&=\frac{1}{8\pi G_n}n_iN_jh^{il}h^{jk}\nabla_ku_l\notag\\
&=-\frac{1}{8\pi G_n}u^iN^j\nabla_jn_i\notag\\
&=-\frac{1}{8\pi G_n}u^\mu N^X\gamma^\nu_{\ \ X}K_{\mu\nu},
\label{condition}
\end{align}
where the third equality in Eq.~(\ref{condition}) follows from $n^\mu u_\mu=0$.

Therefore, the black hole entropy becomes
\begin{align}
S_{\rm BH}=&-\lim_{\epsilon\rightarrow0}(I_{E,\epsilon}-I'_{\infty,E,\epsilon})\notag\\
= &\frac{1}{4G_n}\int_{\Gamma_{E}}d^{n-2}x\sqrt{G}+\frac{1}{4G_{n-1}}\int_{\partial\Gamma_{E}}d^{n-3}x\sqrt{p}\notag\\
= &S_{\rm HEE},
\end{align}
where, in the second equality, we have used the condition (\ref{conical condition BH}) and $S_{\rm HEE}$ is the holographic entanglement entropy in the DGP brane model, given by Eq.~(\ref{HEE EH}).

In Ref.~\cite{Kanai:2026tnt}, which studied the RS II model, a term contributing to the angular momentum at spatial infinity remains on the brane. This term was interpreted as an angular momentum contribution arising from the presence of the brane. In contrast, in the present case, the corresponding contribution arising from the brane boundary term is canceled by the contribution from the brane-localized Einstein-Hilbert term, and thus no such term remains. In the following sections, we extend the gravitational theory on the brane and investigate how this result is modified.

\section{Equality of Gravitational Entropy and Holographic Entanglement Entropy in Lovelock Braneworlds}
\label{sec:lovelock}

In the previous section, we examined the relation between the gravitational entropy and the holographic entanglement entropy of braneworld black holes in the DGP model. We showed that the two entropies coincide for general stationary configurations.

In this section, we extend our analysis by replacing the brane-localized Einstein-Hilbert term in the DGP model with a general Lovelock gravitational action. We investigate the relation between the gravitational entropy and the holographic entanglement entropy in this generalized setup and show that the two entropies also coincide for general stationary braneworld black holes in Lovelock gravity.

The action with Lovelock gravity on the brane is given by
\begin{align}
I=\frac{1}{16\pi G_n}\int_{M_n}d^nx\sqrt{-g}(R-2\Lambda)+\int_{M_{n-1}}d^{n-1}x\sqrt{-q}\left(-\sigma+\frac{1}{16\pi G_{n-1}}\sum_{m}c_m\mathcal{L}_m+\frac{1}{8\pi G_n}[K]^-\right)+I_{\infty},
\label{action lovelock}
\end{align}
where $I_{\infty}$ is the boundary terms at infinity as in the DGP model and $c_m$ are coefficients of Lovelock terms, 
\begin{align}
\mathcal{L}_m=\frac{1}{2^m}q^{\mu_1\nu_1\mu_2\nu_2\cdots \mu_m\nu_m}_{\rho_1\sigma_1\rho_2\sigma_2\cdots \rho_m\sigma_m}{}^{(n-1)}R_{\mu_1\nu_1}^{\ \ \ \rho_1\sigma_1}{}^{(n-1)}R_{\mu_2\nu_2}^{\ \ \ \rho_2\sigma_2}\cdots {}^{(n-1)}R_{\mu_m\nu_m}^{\ \ \ \rho_m\sigma_m},
\end{align}
and $\binom{m-1}{k}$ is the binomial coefficients.The tensor $q^{\mu_1\nu_1\mu_2\nu_2\cdots \mu_m\nu_m}_{\rho_1\sigma_1\rho_2d\sigma_2\cdots \rho_m\sigma_m}$ is defined as
\begin{align}
q^{\mu_1\nu_1\mu_2\nu_2\cdots \mu_m\nu_m}_{\rho_1\sigma_1\rho_2d\sigma_2\cdots \rho_m\sigma_m}&:=(2m)!\ q^{\mu_1}_{\ \ [\rho_1}q^{\nu_1}_{\ \ \sigma_1}q^{\mu_2}_{\ \ \rho_2}q^{\nu_2}_{\ \ \sigma_2}\cdots q^{\mu_m}_{\ \ \rho_{m}}q^{\nu_m}_{\ \ \sigma_m]}.
\end{align}
Here, the brackets $[\ ]$ denote complete antisymmetrization over the enclosed indices.

The variation of the action is
\begin{align}
\delta I=&\frac{1}{16\pi G_n}\int_{M_n} d^nx\sqrt{-g}\left(R_{MN}-\frac{1}{2}g_{MN}R+\Lambda g_{MN}\right)\notag\\
&+\frac{1}{16\pi G_{n-1}}\int_{M_{n-1}} d^{n-1}x\sqrt{-q}\left(\sum_{m}c_mE_{\mu\nu}^{(m)}-8\pi G_{n-1}{}^{(n-1)}T_{\mu\nu}\right),
\end{align}
where
\begin{align}
E_{\mu\nu}^{(m)}=\frac{m}{2^m}q^{\mu_1\nu_1\mu_2\nu_2\cdots \mu_m\nu_m}_{\rho_1\sigma_1\rho_2d\sigma_2\cdots \rho_m\sigma_m}q_{\mu_1\mu}{}^{(n-1)}R_{\nu\nu_1}^{\ \ \ \rho_1\sigma_1}{}^{(n-1)}R_{\mu_2\nu_2}^{\ \ \ \rho_2\sigma_2}\cdots {}^{(n-1)}R_{\mu_m\nu_m}^{\ \ \ \rho_m\sigma_m}-\frac{1}{2}\mathcal{L}_m q_{\mu\nu}.
\end{align}
We first derive the effective gravitational equation on the brane by varying the DGP action with respect to the bulk and induced metrics. The variation with respect to the bulk metric yields the $n$-dimensional vacuum Einstein equation,
\begin{align}
G_{MN}+\Lambda g_{MN}=0,
\end{align}
while the variation with respect to the induced metric gives the brane equation
\begin{align}
\sum_mc_mE_{\mu\nu}^{(m)}=8\pi G_{n-1}{}^{(n-1)}T_{\mu\nu},
\label{EOM brain}
\end{align}
where ${}^{(n-1)}T_{\mu\nu}$ is the energy momentum tensor on the brane as given in Eq.~(\ref{T}).

We will show that the black hole entropy obtained from the Euclidean action of Lovelock gravity localized on the brane coincides with the holographic entanglement entropy derived from the replica formula, Eq.~(\ref{replica formula}), for general static braneworld spacetimes, while the two entropies do not generally coincide in stationary braneworld spacetimes. For the brane Lovelock theory, the holographic entanglement entropy obtained from the replica formula takes the form
\begin{align}
S_{\rm HEE}=&\ \frac{1}{4G_n}\int_{\Gamma_E}d^{n-2}x\sqrt{G}+\frac{1}{4G_{n-1}}\int_{\partial\Gamma_E}d^{n-3}x\sqrt{p}\sum_m\frac{m}{2^{m-1}}c_mp^{A_1B_1\cdots A_{m-1}B_{m-1}}_{C_1D_1\cdots C_{m-1}D_{m-1}}{}^{(n-3)}R_{A_1B_1}^{\ \ \ C_1D_1}\cdots {}^{(n-3)}R_{A_{m-1}B_{m-1}}^{\ \ \ C_{m-1}D_{m-1}},
\label{HEE lovelock}
\end{align}
where the tensor $p^{A_1B_1\cdots A_{m-1}B_{m-1}}_{C_1D_1\cdots C_{m-1}D_{m-1}}$ is defined as
\begin{align}
p^{A_1B_1\cdots A_{m-1}B_{m-1}}_{C_1D_1\cdots C_{m-1}D_{m-1}}:=(2m-2)!\ p^{A_1}_{\ \ [C_1}p^{B_1}_{\ \ D_1}\cdots p^{A_{m-1}}_{\ \ C_{m-1}}p^{B_{m-1}}_{\ \ D_{m-1}]}.
\end{align}

We first calculate the gravitational entropy in Lovelock gravity. To this end, we rewrite the Lovelock gravitational action in canonical form, which allows us to extract the entropy from the horizon contribution. The resulting canonical action is given by
\begin{align}
I=&\int_{M_n}d^{n}x \left(\pi^{ij}\dot{h}_{ij}-N\mathcal{H}-N^i\mathcal{H}_i-\frac{\sqrt{-g}}{8\pi G_n}\nabla_M(u^N\nabla_Nu^M-u^MH)-2\sqrt{h}\bar{\nabla}_i\left(\frac{N_j\pi^{ij}}{\sqrt{h}}\right)\right)\notag\\
&+\int_{M_{n-1}}d^{n-1}x\sqrt{-q}\Biggl(\frac{\check{\pi}^{XY}}{\sqrt{-q}}\dot{\gamma}_{XY}-\sigma+\frac{1}{8\pi G_n}\left[K\right]^--\frac{1}{\sqrt{\gamma}}\check{\mathcal{H}}-\frac{N^X}{\sqrt{-q}}\check{\mathcal{H}}_X\notag\\
&-\frac{1}{16\pi G_n}D_{\rho}\Biggl(\sum_m\frac{4m}{2^m}(2m-1)!\ c_m\gamma^{\mu_1}_{\ \ [\rho_1}\gamma^{\nu_1}_{\ \ \sigma_1}\cdots \gamma^{\nu_{m-1}}_{\ \ \sigma_{m-1}}\gamma^{\mu_m}_{\ \ \rho_m]}\gamma^{\mu}_{\ \ \mu_1}\gamma^{\nu \rho_1}\left(q^{\rho}_{\ \ \mu}u^\sigma D_\sigma u_\nu-u^\rho D_\mu u_\nu\right) \notag\\
&\qquad\qquad\qquad\qquad\times {}^{(n-1)}R_{\nu_1\mu_2}^{\ \ \ \sigma_1\rho_2}\cdots {}^{(n-1)}R_{\nu_{m-1}\mu_m}^{\ \ \ \sigma_{m-1}\rho_m}\Biggr)+\cdots\Biggr)+I_{\infty},
\end{align}
where $\check{\pi}_{XY}$ is defined as
\begin{align}
\check{\pi}^X_{\ \ Y}=-\frac{1}{56\pi G_{n-1}}\sqrt{-q}&\sum_m\frac{(2m)!}{2^m}c_m\sum_s^{m-1}\frac{(-4)^{m-s}}{s![2(m-s)-1]!!}\gamma^{X_1Y_1\cdots X_sY_s\cdots X_{m}X}_{Z_1W_1\cdots Z_sW_s\cdots Z_{m}Y}\notag\\
&\times{}^{(n-2)}R_{X_1Y_1}^{\ \ \ Z_1W_1}\cdots{}^{(n-2)}R_{X_sY_s}^{\ \ \ Z_sY_s}{}^{(n-2)}H_{X_{s+1}}^{\ \ Z_{s+1}}{}^{(n-2)}H_{Y_{s+1}}^{\ \ W_{s+1}}\cdots{}^{(n-2)}H_{X_{m}}^{\ \ Z_{m}},
\end{align}
while $\check{\mathcal{H}}$ and $\check{\mathcal{H}}_X$ take the forms
\begin{align}
\check{\mathcal{H}}=&-\frac{\sqrt{\gamma}}{8\pi G_{n-1}}\sum_{m}c_mu^\mu u^\nu E_{\mu\nu}^{(m)}\notag\\
=&-\frac{1}{16\pi G_{n-1}}\sqrt{-q}\sum_m\frac{c_m}{2^m}\gamma^{X_1Y_1\cdots X_{m}Y_{m}}_{Z_1W_1\cdots Z_{m}W_{m}}{}^{(n-2)}R_{X_1Y_1}^{\ \ \ Z_1W_1}\cdots {}^{(n-2)}R_{X_mY_m}^{\ \ \ Z_mW_m},\\
\check{\mathcal{H}}_X=&-\frac{\sqrt{\gamma}}{8\pi G_{n-1}}\sum_mc_mu^\mu\gamma^\nu_{\ X}E_{\mu\nu}^{(m)}\notag\\
=&-2\gamma_{XY}\tilde{\nabla}_Z\check{\pi}^{YZ}.
\end{align}
Here, the tensor $\gamma^{X_1Y_1\cdots X_{m}Y_{m}}_{Z_1W_1\cdots Z_{m}W_{m}}$ are defined as
\begin{align}
\gamma^{X_1Y_1\cdots X_{m}Y_{m}}_{Z_1W_1\cdots Z_{m}W_{m}}:=(2m)!\ \gamma^{X_1}_{\ \ [Z_1}\gamma^{Y_1}_{\ \ W_1}\cdots \gamma^{X_{m}}_{\ \ Z_{m}}\gamma^{Y_{m}}_{\ \ W_{m}]}.
\end{align}
This result was obtained in Ref.~\cite{Teitelboim:1987zz}. Since we choose the time coordinate such that the shift vector vanishes on the horizon, the total derivative term on the brane involving the shift vector vanishes in the present calculation. We therefore do not write it explicitly. The total derivative terms in Lovelock gravity that do not involve time derivatives or the shift vector are given by the expression above, as can be seen in Ref.~\cite{Kanai:2026tnt}.

Therefore, the Euclidean action for the spacetime with a cutoff at the lapse function $N=\epsilon$ is given by
\begin{align}
I_{\rm E,\epsilon}=&\beta_0\int_{\Sigma_{n,E,\epsilon}}d^{n-1}x \left(-i\pi^{ij}\dot{h}_{ij}+N\mathcal{H}+N^i\mathcal{H}_i\right)\notag\\
&+\beta_0\int_{\Sigma_{n-1,E,\epsilon}}d^{n-2}x\sqrt{q}\Biggl(-i\frac{\check{\pi}^{XY}}{\sqrt{q}}\dot{\gamma}_{XY}+\sigma-\frac{1}{8\pi G_n}\left[{}^{(n-2)}K\right]^-+\frac{1}{\sqrt{\gamma}}\check{\mathcal{H}}+\frac{N^X}{\sqrt{q}}\check{\mathcal{H}}_X+\frac{2n_iN_j\pi^{ij}}{N\sqrt{h}}\Biggr)\notag\\
&-\frac{\beta_0}{8\pi G_n}\int_{\Gamma_{E,\epsilon}}d^{n-2}x\sqrt{G}r^i\partial_iN\notag\\
&-\frac{\beta_0}{8\pi G_{n-1}}\int_{\partial\Gamma_{E,\epsilon}}d^{n-3}x\sqrt{p}\sum_m\frac{m}{2^{m-1}}c_m p^{A_1B_1\cdots A_{m-1}B_{m-1}}_{C_1D_1\cdots C_{m-1}D_{m-1}}r^X\partial_XN{}^{(n-3)}R_{A_1B_1}^{\ \ \ C_1D_1}\cdots {}^{(n-3)}R_{X_{m-1}Y_{m-1}}^{\ \ \ C_{m-1}D_{m-1}}+I'_{\infty,E,\epsilon}\notag\\
=&\frac{\beta_0}{8\pi G_{n}}\int_{\Sigma_{n-1,E,\epsilon}} d^{n-2}x\sqrt{\gamma}\left([K_{\mu\nu}u^\mu \gamma^\nu_{\ X}N^X]^-+\frac{2n_iN_j\pi^{ij}}{\sqrt{h}}\right)-\frac{\beta_0}{8\pi G_n}\int_{\Gamma_{E,\epsilon}}d^{n-2}x\sqrt{G}r^i\partial_iN\notag\\
&-\frac{\beta_0}{8\pi G_{n-1}}\int_{\partial\Gamma_{E,\epsilon}}d^{n-3}x\sqrt{p}\sum_m\frac{m}{2^{m-1}}c_m p^{A_1B_1\cdots A_{m-1}B_{m-1}}_{C_1D_1\cdots C_{m-1}D_{m-1}}r^X\partial_XN{}^{(n-3)}R_{A_1B_1}^{\ \ \ C_1D_1}\cdots {}^{(n-3)}R_{X_{m-1}Y_{m-1}}^{\ \ \ C_{m-1}D_{m-1}}+I'_{\infty,E,\epsilon},
\end{align}
where, in the second equality, we have used the stationarity of the spacetime and the Hamiltonian and momentum constraints, $\dot{h}_{ij}=\mathcal{H}=\mathcal{H}_i=0$, as in the previous section. We have also neglected the contributions from terms proportional to the shift vector, since the shift vector vanishes on $\Gamma_{E,\epsilon}$ in the limit $\epsilon\to 0$, and these contributions therefore vanish in this limit on both $\Gamma_{E,\epsilon}$ and $\partial\Gamma_{E,\epsilon}$. Meanwhile, the terms on $\Sigma_{n-1,\epsilon}$ have been rearranged using the stationarity of the spacetime and the relations obtained from the $(u,u)$ and $(u,i)$ components of the brane equations of motion, as given below:
\begin{align}
\dot{\gamma}_{XY}&=0,\\
\frac{1}{\sqrt{\gamma}}\check{\mathcal{H}}+\sigma-\frac{1}{8\pi G_n}\left[{}^{(n-2)}K\right]^-&=0,\\
N^X\check{\mathcal{H}}_X&=\frac{1}{8\pi G_n}\sqrt{\gamma}[K_{\mu\nu}u^\mu\gamma^\nu_{\ X}N^X]^-.
\label{lovelock constraint}
\end{align}
It should be noted that Euclideanization changes the signs of some terms, while leaving those of others unchanged. Therefore, as in the previous section, the contribution associated with the angular momentum at bulk infinity is canceled by the Lovelock constraint on the brane, given by Eq.~(\ref{lovelock constraint}).

Then, the black hole entropy is
\begin{align}
S_{\rm BH}=&-\lim_{\epsilon\rightarrow0}(I_{\rm E,\epsilon}-I'_{\infty,E,\epsilon})\notag\\
=&\ \frac{1}{4G_n}\int_{\Gamma_{E}}d^{n-2}x\sqrt{G}+\frac{1}{4G_{n-1}}\int_{\partial\Gamma_{E}}d^{n-3}x\sqrt{p}\sum_m\frac{m}{2^{m-1}}c_m p^{A_1B_1\cdots A_{m-1}B_{m-1}}_{C_1D_1\cdots C_{m-1}D_{m-1}}{}^{(n-3)}R_{A_1B_1}^{\ \ \ C_1D_1}\cdots {}^{(n-3)}R_{X_{m-1}Y_{m-1}}^{\ \ \ C_{m-1}D_{m-1}}\notag\\
=&\ S_{\rm HEE}
\end{align}
where, in the third equality, we have used the decomposition of both the first and second terms into components along the direction of the unit normal vector $r_{\mu}$ to the horizon and those orthogonal to it, together with the fact that the extrinsic curvature associated with $r_{\mu}$ vanishes on the horizon and $S_{\rm HEE}$ is the holographic entanglement entropy with lovelock gravity in brane, given by Eq~(\ref{HEE lovelock}).

Therefore, we have shown that, as in the DGP model discussed in the previous section, the gravitational entropy coincides with the holographic entanglement entropy for general stationary braneworld spacetimes with Lovelock gravity on the brane.

\section{Equality of gravitational entropy and holographic entanglement entropy in $f(R)$ braneworlds}
\label{sec:fr}

We next consider $f(R)$ gravity localized on the brane and investigate whether the equality between gravitational entropy and holographic entanglement entropy, established in the previous sections, also holds for general stationary spacetimes. Lovelock gravity is a higher-curvature theory whose equations of motion nevertheless remain of second order. It is therefore natural to extend our analysis to theories with genuinely higher-order equations of motion. As a representative example, we consider $f(R)$ gravity, whose equations of motion contain higher derivatives of the metric. The higher-derivative degrees of freedom in $f(R)$ gravity can be reformulated through an equivalent Lagrange-multiplier formulation by introducing an auxiliary scalar field in Ref.~\cite{Deruelle:2009pu}. In this section, we employ this formulation to investigate the relation between gravitational entropy and holographic entanglement entropy in $f(R)$ gravity localized on the brane, following the same approach as in the previous sections.

The action with $f(R)$ gravity located on the brane is 
\begin{align}
I=\frac{1}{16\pi G_n}\int_{M_n}d^nx\sqrt{-g}(R-2\Lambda)+\int_{M_{n-1}}d^{n-1}x\sqrt{-q}\left(-\sigma+\frac{1}{2}f\left({}^{(n-1)}R\right)+\frac{1}{8\pi G_n}[K]^-\right)+I_{\infty}.
\label{action fR}
\end{align}
We normalize $f(R)$ such that the Einstein-Hilbert limit is recovered for $f(R)=R/(8\pi G_{n-1})$. To analyze this theory, it is convenient to rewrite the brane action in a scalar-tensor form by introducing an auxiliary field $s$. We consider
\begin{align}
I=\frac{1}{16\pi G_n}\int_{M_n}d^nx\sqrt{-g}(R-2\Lambda)+\int_{M_{n-1}}d^{n-1}x\sqrt{-q}\left(-\sigma+\frac{1}{2}f(s)-\frac{1}{2}\phi\left(s-{}^{(n-1)}R\right)+\frac{1}{8\pi G_n}[K]^-\right)+I_{\infty},
\end{align}
where $\phi$ is a Lagrange multiplier.

We first derive the effective gravitational equation on the brane by varying the $f(R)$ brane action with respect to the bulk and induced metrics. The variation with respect to the bulk metric yields the $n$-dimensional vacuum Einstein equation,
\begin{align}
G_{MN}+\Lambda g_{MN}=0,
\end{align}
while the variation with respect to the induced metric gives the brane equation
\begin{align}
\phi{}^{(n-1)}G_{\mu\nu}&=8\pi G_{n-1}{}^{(n-1)}T_{\mu\nu},\\
\phi&=f'(s),\\
s&={}^{(n-1)}R,
\end{align}
where $f'(s)$ is defined as
\begin{align}
f'(s):=\frac{df}{ds}(s),
\end{align}
and, as in the preceding sections, ${}^{(n-1)}T_{\mu\nu}$ is defined by Eq.~(\ref{T}).

We will show that the black hole entropy obtained from the Euclidean action of $f(R)$ gravity located on the brane coincides with the holographic entanglement entropy derived from the replica formula, Eq.~(\ref{replica formula}). For general static braneworld spacetimes with $f(R)$ gravity on the brane, the resulting holographic entanglement entropy is given by
\begin{align}
S_{\rm HEE}=&\ \frac{1}{4G_n}\int_{\Gamma_E}d^{n-2}x\sqrt{G}+2\pi\int_{\partial\Gamma_E}d^{n-3}x\sqrt{p}f'\left({}^{(n-1)}R\right).
\label{HEE fr}
\end{align}

To compute the black hole entropy, we first write the gravitational action in the canonical form. In this formulation, the action is decomposed into the bulk Hamiltonian and boundary contributions, allowing the entropy to be extracted from the boundary term associated with the black hole horizon.
\begin{align}
I=&\int_{M_n}d^{n}x \left(\pi^{ij}\dot{h}_{ij}-N\mathcal{H}-N^i\mathcal{H}_i-\frac{\sqrt{-g}}{8\pi G_n}\nabla_M(u^N\nabla_Nu^M-u^MH)-2\sqrt{h}\bar{\nabla}_i\left(\frac{N_j\pi^{ij}}{\sqrt{h}}\right)\right)\notag\\
&+\int_{M_{n-1}}d^{n-1}x\Bigl(\breve{\pi}^{XY}\dot{\gamma}_{XY}+\lambda\dot{\phi}-\sigma-N\breve{\mathcal{H}}-N^X\breve{\mathcal{H}}_X+\frac{1}{8\pi G_n}[K]^-\notag\\
&\qquad\qquad\qquad\qquad+\partial_{\mu}(\sqrt{-q}\phi{}^{(n-2)}H u^\mu)-\partial_X(\sqrt{\gamma}\phi\gamma^{XY}\partial_YN+2\breve{\pi}^{XY}N_Y-\sqrt{\gamma}\partial^X\phi N)\Bigr)+I_{\infty},
\end{align}
where $\breve{\pi}_{XY}$ and $\lambda$ are, respectively, defined as
\begin{align}
\breve{\pi}_{XY}&=\frac{\sqrt{\gamma}}{2}\left[\phi\left({}^{(n-2)}H_{XY}-\frac{1}{n-1}\gamma_{XY}{}^{(n-2)}H\right)-\frac{\gamma_{XY}}{N}(\dot{\phi}-N^X\partial_X\phi)\right],\\
\lambda&=-\sqrt{\gamma}{}^{(n-2)}H.
\end{align}
Moreover, $\breve{\mathcal{H}}$ and $\breve{\mathcal{H}}_X$ are given by
\begin{align}
\breve{\mathcal{H}}=&\frac{2}{\sqrt{\gamma}}\left(\frac{\breve{\pi}_{XY}\breve{\pi}^{XY}-\frac{1}{n-1}\breve{\pi}^2}{\phi}-\frac{n-2}{4(n-1)}\lambda^2-\frac{1}{n-1}\breve{\pi}\lambda\right)+\frac{1}{2}(\phi s-f(s)-\phi{}^{(n-2)}R+2\tilde{\nabla}_X\tilde{\nabla}^X\phi),\notag\\
\breve{\mathcal{H}}_X=&-2\sqrt{\gamma}\tilde{\nabla}_Y\left(\frac{\breve{\pi}^Y_{\ X}}{\sqrt{\gamma}}\right)+\lambda\tilde{\nabla}_X\phi.
\end{align}

Therefore, the Euclidean action becomes
\begin{align}
I_{E,\epsilon}=&\frac{1}{16\pi G_n}\int_{M_{n,E,\epsilon}}d^nx\left(-i\pi^{ij}\dot{h}_{ij}+N\mathcal{H}+N^i\mathcal{H}_i+2\partial_{M}(\sqrt{-g}H u^M)+2\partial_i(\sqrt{h}h^{ij}\partial_jN)\right)\notag\\
&+\int_{M_{n-1,E,\epsilon}}d^{n-1}x\Bigl(-i\breve{\pi}^{XY}\dot{\gamma}_{XY}-i\lambda\dot{\phi}+\sigma+N\breve{\mathcal{H}}+N^X\breve{\mathcal{H}}_X-\frac{1}{8\pi G_n}[K]^-+\sqrt{\gamma}\frac{2n_iN_j\pi^{ij}}{\sqrt{h}}\notag\\
&\qquad\qquad\qquad\qquad-\partial_{\mu}(\sqrt{q}\phi{}^{(n-2)}H u^\mu)+\partial_X(\sqrt{\gamma}\phi\gamma^{XY}\partial_YN+2\breve{\pi}^{XY}N_Y+\sqrt{\gamma}\partial^X\phi N)\Bigr)+I'_{\infty,E,\epsilon}\notag\\
=&\ \frac{\beta_0}{8\pi G_{n}}\int_{\Sigma_{n-1,E,\epsilon}} d^{n-2}x\sqrt{\gamma}\left([K_{\mu\nu}u^\mu \gamma^\nu_{\ X}N^X]^-+\frac{2n_iN_j\pi^{ij}}{\sqrt{h}}\right)-\frac{\beta_0}{8\pi G_n}\int_{\Gamma_{E,\epsilon}} d^{n-2}x\sqrt{G}r^i\partial_iN\notag\\
&\ -\beta_0\int_{\partial\Gamma_{E,\epsilon}}d^{n-3}x \sqrt{p}r_X\left(\phi\gamma^{XY}\partial_YN+\partial^X\phi N\right)+I'_{\infty,E,\epsilon},
\end{align}
where, in the second equality, we have used the stationarity of the spacetime, the condition $r_\mu u^\mu=0$, and the Hamiltonian and momentum constraints for the full system on the brane, corresponding to the $(u,u)$ and $(u,X)$ components of the equations of motion, respectively. It should be noted that the Wick rotation does not affect all terms in the same way. Depending on their dependence on the time coordinate, some terms acquire an additional minus sign, whereas others remain unchanged. As in the previous sections, the terms on the brane cancel those from the bulk and hence vanish. We have also omitted terms involving the shift vector, since the latter vanishes on the horizon in the limit $\epsilon\to 0$.

Therefore, the black hole entropy with $f(R)$ gravity located on the brane become
\begin{align}
S_{\rm BH}=&\ -\lim_{\epsilon\rightarrow0}(I_{E,\epsilon}-I'_{\infty,E,\epsilon})\notag\\
=&\  \lim_{\epsilon\rightarrow0}\left[\frac{\beta_0}{8\pi G_n}\int_{\Gamma_{E,\epsilon}} d^{n-2}x\sqrt{G}r^i\partial_iN+\beta_0\int_{\partial\Gamma_{E,\epsilon}}d^{n-3}x \sqrt{p}r_X\phi\gamma^{XY}\partial_YN\right]\notag\\
=&\ \frac{1}{4G_n}\int_{\Gamma_E} d^{n-2}x\sqrt{G}+\int_{\partial\Gamma_E}d^{n-3}x \sqrt{p}f'\left({}^{(n-1)}R\right)\notag\\
=&\ S_{\rm HEE},
\end{align}
where we have used $N=N^i=0$ on the horizon in the second equality and the condition for the absence of a conical singularity, Eq.~(\ref{conical condition BH}), in the third equality. The quantity $S_{\rm HEE}$ denotes the holographic entanglement entropy for $f(R)$ gravity localized on the brane and is given by Eq.~(\ref{HEE fr}).

As in the previous cases, gravitational entropy and holographic entanglement entropy coincide for general stationary spacetimes in $f(R)$ gravity localized on the brane.

\section{Conclusion and Discussion}
\label{sec:conclusion}

In this work, we have investigated the relation between gravitational entropy and holographic entanglement entropy for braneworld black holes with gravitational dynamics localized on the brane. We first considered the DGP brane model, in which an Einstein--Hilbert term is included in the brane action, and compared the black hole entropy obtained from the Euclidean gravitational action with the holographic entanglement entropy. We demonstrated that the two entropy prescriptions coincide for general stationary braneworld spacetimes, without assuming any specific form of the spacetime metric, relying only on the stationarity of the spacetime.

We then generalized the brane-localized Einstein-Hilbert term to Lovelock gravity. By extending the analysis to higher-curvature gravitational theories whose equations of motion remain of second order, we showed that the gravitational entropy and holographic entanglement entropy also coincide for general stationary spacetimes. Thus, the correspondence established in the DGP model persists in the presence of higher-curvature corrections described by Lovelock gravity.

We further extended the analysis to $f(R)$ gravity localized on the brane. Since the equations of motion in $f(R)$ gravity contain higher derivatives of the metric, we employed an equivalent Lagrange-multiplier formulation with an auxiliary scalar field to carry out the analysis. We again found that gravitational entropy agrees with holographic entanglement entropy for general stationary spacetimes. This result demonstrates that the correspondence also holds in theories with genuinely higher-order equations of motion.

Taken together, our results show that gravitational entropy and holographic entanglement entropy coincide for general stationary braneworld spacetimes in all the gravitational theories considered in this work: the DGP model, Lovelock gravity, and $f(R)$ gravity. The persistence of this agreement across these different classes of theories suggests that the correspondence is not restricted to a particular form of the gravitational action localized on the brane. Rather, it appears to reflect a more general feature of braneworld gravity.

There remain several interesting directions for future investigation. It would be important to extend the present analysis to other classes of brane-localized gravitational theories, particularly more general higher-curvature theories with higher-order equations of motion. It would also be interesting to investigate the correspondence in more general braneworld settings and to clarify its geometric and physical origin. Such investigations may further deepen our understanding of the relation between gravitational entropy and holographic entanglement entropy in braneworld gravity.

\bibliography{references}

\end{document}